\documentclass[a4paper,fleqn]{cas-dc}

\usepackage[numbers]{natbib}
\usepackage{booktabs}
\usepackage{placeins} 
\usepackage{orcidlink}
\usepackage{xcolor}

\protected\def\safeorcid#1{\orcidlink{#1}}

\def\tsc#1{\csdef{#1}{\textsc{\lowercase{#1}}\xspace}}
\tsc{WGM}
\tsc{QE}
\tsc{EP}
\tsc{PMS}
\tsc{BEC}
\tsc{DE}

\begin{document}

\let\WriteBookmarks\relax
\def\floatpagepagefraction{1}
\def\textpagefraction{.001}

\ExplSyntaxOn
\cs_gset:Npn \__first_footerline:
  { \group_begin: \small \sffamily \__short_authors: \group_end: }
\ExplSyntaxOff

\shorttitle{Vacancy and Helium-Cluster Thermomigration in Tungsten}

\shortauthors{S Alturk et~al.}

\title[mode=title]
{Vacancy and Helium-Cluster Thermomigration in Tungsten}

\author[1]{Sanad Alturk\,\safeorcid{0009-0008-8448-0983}}

\author[1,2]{Enrique Martinez\,\safeorcid{0000-0002-2690-2622}}
\cormark[1]
\ead{enrique@clemson.edu}

\credit{Conceptualization of this study, Methodology, Software}

\affiliation[1]{
  organization={School of Mechanical and Automotive Engineering,
  Clemson University},
  city={Clemson},
  state={SC},
  postcode={29364},
  country={USA}
}

\affiliation[2]{
  organization={Department of Materials Science and Engineering,
  Clemson University},
  city={Clemson},
  state={SC},
  postcode={29364},
  country={USA}
}

\cortext[cor1]{Corresponding author}
\begin{abstract}
Strong thermal gradients in tungsten plasma-facing components can bias the migration of irradiation-induced defects and thereby alter their redistribution, interaction, and retention. Here, we extend an atomistic jump-statistics framework previously developed for hydrogen to quantify the thermomigration of a single vacancy and interstitial He$_1$--He$_4$ clusters in body-centered-cubic tungsten. Non-equilibrium molecular-dynamics simulations performed over several thermal gradients are used to determine the directional jump ratio, while geometrically derived projected jump lengths convert the fitted directional-bias coefficient into the heat-of-transport coefficient $Q^*=\alpha kT^2$, where $k$ is the Boltzmann constant, T the temperature and $\alpha$ a constant. All investigated defects exhibit negative values of $\alpha$, corresponding to net migration toward the hot region. He$_4$ has the largest response, $\alpha=-2.2457\times10^{-2}$~K$^{-1}$, followed closely by the vacancy, whereas He$_3$ exhibits the weakest response despite its intermediate cluster size. Independent constant-temperature simulations over 1000--1600~K show that He$_3$ is also the least mobile cluster and has the largest effective migration energy, $E_m^{\mathrm{eff}}=0.271$~eV. In contrast, He$_4$ is more mobile than He$_3$ and has the smallest effective migration energy, 0.087~eV. Agreement between the He$_3$ concentration profile predicted from the jump statistics and that obtained from a long molecular dynamics trajectory provides an independent validation of the method. These results show that thermomigration cannot be inferred from cluster size, diffusivity, or migration energy alone and provide defect-specific heat-of-transport parameters for higher-length-scale models of tungsten under fusion-relevant thermal gradients.
\end{abstract}


\maketitle

\section{Introduction}

One of the principal challenges in realizing fusion energy as a practical and sustainable power source is ensuring the long-term reliability of plasma-facing components (PFCs) \cite{Knaster2016}. These materials are continuously exposed to extreme operating conditions, including intense particle bombardment, neutron irradiation, and heat fluxes that can exceed tens of MW\,m$^{-2}$, producing large thermal gradients across divertor and first wall (FW) components \cite{Pitts2019,Hirai2016}. Tungsten (W) is the primary candidate material for future divertors and FW because of its high melting temperature, excellent thermal conductivity, low sputtering yield, and relatively low tritium retention \cite{Hirai2016,Philipps2011,Roth2011}. 
Nevertheless, the harsh fusion environment progressively modifies its microstructure, ultimately affecting both material integrity and reactor performance \cite{Rieth2013}. Beyond pure tungsten, W-based refractory high-entropy alloys such as WTaCrV have demonstrated strong irradiation resistance, while compositional control has been predicted to improve their intrinsic ductility \cite{ElAtwani2019,Korpe2025}. Recent rate-theory and molecular dynamics (MD) analyses further show that defect diffusivities and effective migration energies strongly influence vacancy--interstitial recombination and the retained defect population under irradiation \cite{Nahavandian2025}. Consequently, understanding the atomistic mechanisms governing defect transport in W under reactor-relevant conditions is essential for predicting the long-term evolution of plasma-facing materials and improving their lifetime \cite{Knaster2016,Nordlund2018}.

Neutron irradiation continuously generates large populations of point defects, primarily vacancies and self-interstitial atoms, and can also produce helium (He) through transmutation reactions within the W lattice \cite{Rieth2013,Nordlund2018,Stoller2012}. In plasma-facing regions, however, the majority of He originates from implantation from the plasma. The subsequent migration and interaction of these irradiation-induced defects and implanted or transmutation-produced He govern numerous degradation mechanisms, including void swelling, irradiation hardening, bubble nucleation and growth, grain-boundary segregation, and surface morphology evolution \cite{Becquart2009,Sefta2013,Sand2015}. In particular, He atoms readily aggregate into small mobile clusters that eventually become trapped by vacancies, forming He--vacancy complexes that act as precursors for nanoscale bubbles \cite{Perez2014,Perez2017,Boisse2014}. These microstructural changes significantly influence the mechanical properties, thermal conductivity, and tritium retention of W, making the transport behavior of vacancies and He clusters a subject of considerable interest in fusion materials research \cite{Hu2014JAP,Hu2015,Hammond2015}.

Although defect diffusion under homogeneous temperature conditions has been extensively investigated using density functional theory, MD, and kinetic Monte Carlo simulations for vacancies, He atoms, and He clusters in W \cite{Becquart2009,Perez2014,Perez2017,Henriksson2006}, considerably less attention has been devoted to transport under thermal gradients. In fusion reactors, however, strong temperature gradients naturally arise across plasma-facing components during steady-state operation and transient events, providing an additional driving force for defect migration through the Soret effect, or thermomigration \cite{Schottky1963,Longhurst1985}. Previous computational studies have demonstrated that thermal gradients can influence the transport of hydrogen and He-related defects in W, suggesting that thermomigration may contribute to the redistribution of irradiation-induced defects and therefore affect the long-term microstructural evolution of plasma-facing materials \cite{Dasgupta2023,Alturk2026,Martinez2021}.

Recently, we developed an atomistic framework to quantify thermomigration in W by relating the ratio of atomic jumps toward the hot and cold regions directly to the heat of transport, $Q^*$ \cite{Alturk2026}. Using non-equilibrium MD (NEMD) simulations, we demonstrated that this approach accurately describes hydrogen transport under thermal gradients and provides a quantitative link between atomistic jump statistics and continuum transport models \cite{Alturk2026}. This framework offers a computationally efficient alternative for evaluating thermomigration without requiring direct measurement of defect fluxes or concentration gradients. However, hydrogen represents only one mobile species in W, and whether the same methodology can be extended to irradiation-induced defects remains an open question.

Vacancies and He clusters differ fundamentally from hydrogen in both their migration mechanisms and their interactions with the surrounding lattice. While vacancies migrate through successive atomic exchanges, He clusters move collectively and exhibit size-dependent mobility arising from the cooperative motion of multiple He atoms \cite{Perez2014,Perez2017,Hu2015}. These differences raise important questions regarding the applicability of a unified thermomigration framework and the influence of defect size on directional transport under thermal gradients. To date, a systematic comparison of the thermomigration behavior of vacancies and small He clusters in W using a common theoretical and computational framework has not been reported.

In this work, we extend our previously developed jump-statistics framework to investigate the thermomigration of vacancies and He clusters containing one to four atoms in body-centered cubic (BCC) W. Using extensive NEMD simulations comprising hundreds of independent trajectories for each defect and thermal gradient, we quantify the directional jump bias, determine the corresponding heat of transport, and examine how thermomigration evolves with defect size. Independent constant-temperature simulations are additionally used to determine the diffusivities and effective migration energies of He$_1$--He$_4$ and to examine the anomalous response of He$_3$. The resulting framework provides a unified atomistic description of thermally driven defect transport in W and establishes quantitative parameters that can be incorporated into mesoscale and continuum models of microstructural evolution in fusion plasma-facing materials.
\section{Methodology}

\subsection{Molecular dynamics simulations}

Non-equilibrium molecular-dynamics simulations were
performed using the Large-scale Atomic/Molecular Massively Parallel
Simulator (LAMMPS)~\cite{Plimpton1995} to investigate the
thermomigration of vacancies and He clusters in BCC W.
Periodic boundary conditions were applied in all three directions, with
the thermal gradient imposed along the $z$ direction.

The W--W interactions were described using the
embedded-atom method potential developed by Marinica \textit{et
al.}~\cite{Marinica2013}. For the He-cluster simulations, the W--He
interaction was represented by the potential developed by Juslin and
Wirth~\cite{Juslin2013}, and the He--He interaction by the modified
Beck pair potential~\cite{Beck1968}.

A thermal gradient was established using Langevin
thermostats applied to a hot slab at the center of the simulation cell
and cold slabs at the periodic cell boundaries. This arrangement
produced two equivalent gradient regions, with the direction toward the
hot region reversing across the cell center. Atoms outside the
thermostatted slabs evolved in the microcanonical (NVE) ensemble, and
the Langevin damping time was 0.1~ps. The gradient magnitude was
calculated as $|\nabla T|=2\Delta T/L_z$, where
$\Delta T=T_{\mathrm{hot}}-T_{\mathrm{cold}}$ and $L_z$ is the cell
length along $z$. 
For both defects, trajectories were recorded only during the production stage, following energy minimization and equilibration under the described thermal conditions.

The vacancy was created by removing one W atom from the perfect
lattice. The He atoms were inserted in compact initial
configurations. Each system was energy minimized with anisotropic box
relaxation to zero pressure before the temperature-controlled
simulation.

Different simulation cells and thermodynamic conditions were employed
for the two defect classes to account for their distinct migration
kinetics. The simulation cells were oriented such that the $x$, $y$, and $z$
directions corresponded to the $[100]$, $[010]$, and $[001]$
crystallographic directions, respectively. Helium-cluster simulations
were performed in cells measuring approximately
$2.53\times2.53\times19.0~\mathrm{nm}^3$, whereas vacancy simulations
employed cells of approximately
$2.53\times2.53\times56.97~\mathrm{nm}^3$.

For the He-cluster simulations, the average temperature
was maintained at 1300~K. Temperature differences of 0, 200, 400, 600,
and 800~K produced nominal gradient magnitudes of approximately 0,
21.05, 42.11, 63.16, and 84.21~K\,nm$^{-1}$, respectively. Each
simulation was equilibrated for 20,000 steps using a timestep of 0.5~fs,
followed by a production run of $10^6$ steps.

Vacancy simulations were performed at an average temperature
of 2300~K using temperature differences of 0, 200, 400, 600, and
800~K, corresponding to nominal gradient magnitudes of approximately 0,
7.02, 14.04, 21.07, and 28.09~K\,nm$^{-1}$, respectively. Owing to the
lower vacancy mobility, higher average temperature and longer
trajectories were required to obtain sufficient migration events. Each
vacancy production trajectory was evolved for $5.0\times10^7$ steps
using a timestep of 1~fs.

For each gradient condition, 250 independent trajectories
were performed for each He-cluster size and 100 longer trajectories
were performed for the vacancy using independent velocity seeds. A
single directional jump ratio was calculated for each trajectory, and
all trajectory-level ratios were retained as the observations used in
the regression analysis; hot- and cold-directed counts were not pooled
across trajectories. Sampling adequacy was verified from stabilization
of the cumulative mean directional jump ratio with increasing ensemble
size.

\subsection{Thermomigration framework}

In our previous work, we developed a jump-statistics framework that establishes a direct relationship between the directional bias of individual diffusion events and the heat-of-transport for hydrogen in W~\cite{Alturk2026}. Rather than relying on equilibrium concentration profiles, the framework demonstrated that thermomigration can be quantified directly from atomistic trajectories through the statistics of individual migration events. In the present work, we demonstrate that this framework is equally applicable to fundamentally different defect species by extending it to single vacancies and He clusters.

According to irreversible thermodynamics, the particle flux under the combined influence of concentration and temperature gradients is given by

\begin{equation}
j=-D\nabla c-D\frac{cQ^*}{kT^2}\nabla T,
\end{equation}

\noindent where $D$ is the diffusivity, $c$ is the species concentration, $T$ is the local temperature, $k$ is Boltzmann's constant, and $Q^*$ is the heat of transport. In our previous work, this continuum description was reformulated in terms of the statistics of individual migration events by expressing the particle flux through directional jumps toward the hot and cold regions. Since the complete derivation has already been presented in Ref.~\cite{Alturk2026}, only the resulting expressions required for the present work are summarized here.

The directional jump ratio is defined as

\begin{equation}
\phi=\frac{N_{\rm hot}}{N_{\rm cold}},
\label{eq:phi}
\end{equation}

\noindent where $N_{\rm hot}$ and $N_{\rm cold}$ denote the numbers of jumps toward the hot and cold regions, respectively. The corresponding heat-of-transport is given by

\begin{equation}
Q^*=\frac{2kT^2}{\bar b |\nabla T |}\frac{1-\phi}{1+\phi},
\end{equation}

\noindent where $\bar b$ is the characteristic jump length projected in the direction of the thermal gradient determined from the geometry of the corresponding migration pathway. Because the applied thermal gradients represent only a small perturbation to equilibrium diffusion, the first-order approximation

\begin{equation}
Q^*\approx\frac{kT^2}{\bar b|\nabla T |}(1-\phi)
\label{eq:qstar}
\end{equation}

\noindent is used throughout this work. In the hydrogen
study, the coefficient relating $Q^*$ to $kT^2$ showed no statistically
significant dependence on either the mean temperature or the applied
gradient over the investigated conditions~\cite{Alturk2026}.
Accordingly, the temperature dependence is retained explicitly through
$Q^*=\alpha kT^2$, while $\alpha$ is treated as a defect-dependent
coefficient.

\subsection{Defect trajectory analysis}

Defect trajectories were analyzed using the Open Visualization Tool
(OVITO)~\cite{Stukowski2010}. For the vacancy simulations,
common-neighbor analysis was used during the LAMMPS trajectory to
identify W atoms whose local environments were not classified as BCC.
These non-BCC atoms were written to the vacancy trajectory and grouped
at each stored frame. The center of mass $z$ coordinate of the
vacancy associated cluster was used to represent the vacancy position.
For each He cluster, the defect position was represented by the center
of mass of its constituent He atoms. Periodic boundary conditions were
accounted for when evaluating successive vacancy and He cluster
positions. For the vacancy, displacements smaller than 0.8~\AA\ were
attributed to thermal vibrations and excluded from the jump statistics.
Displacements associated with thermal vibrations were likewise excluded
before identifying He-cluster migration events.

Each accepted event was classified as directed toward either the
central hot slab or the cold slabs at the cell boundaries. At zero
gradient, the same geometrical labels were retained, although the two
directions are thermodynamically equivalent and therefore require
$\phi\simeq1$. For every trajectory, one value
$\phi_i=N_{{\rm hot},i}/N_{{\rm cold},i}$ was extracted. Linear
regression was then performed using all trajectory-level values.

\begin{figure*}[!t]
    \centering


    \begin{minipage}[t]{0.27\textwidth}
        \centering
        \includegraphics[
            width=0.94\linewidth,
            trim=0.15cm 0.3cm 0.15cm 0.3cm,
            clip
        ]{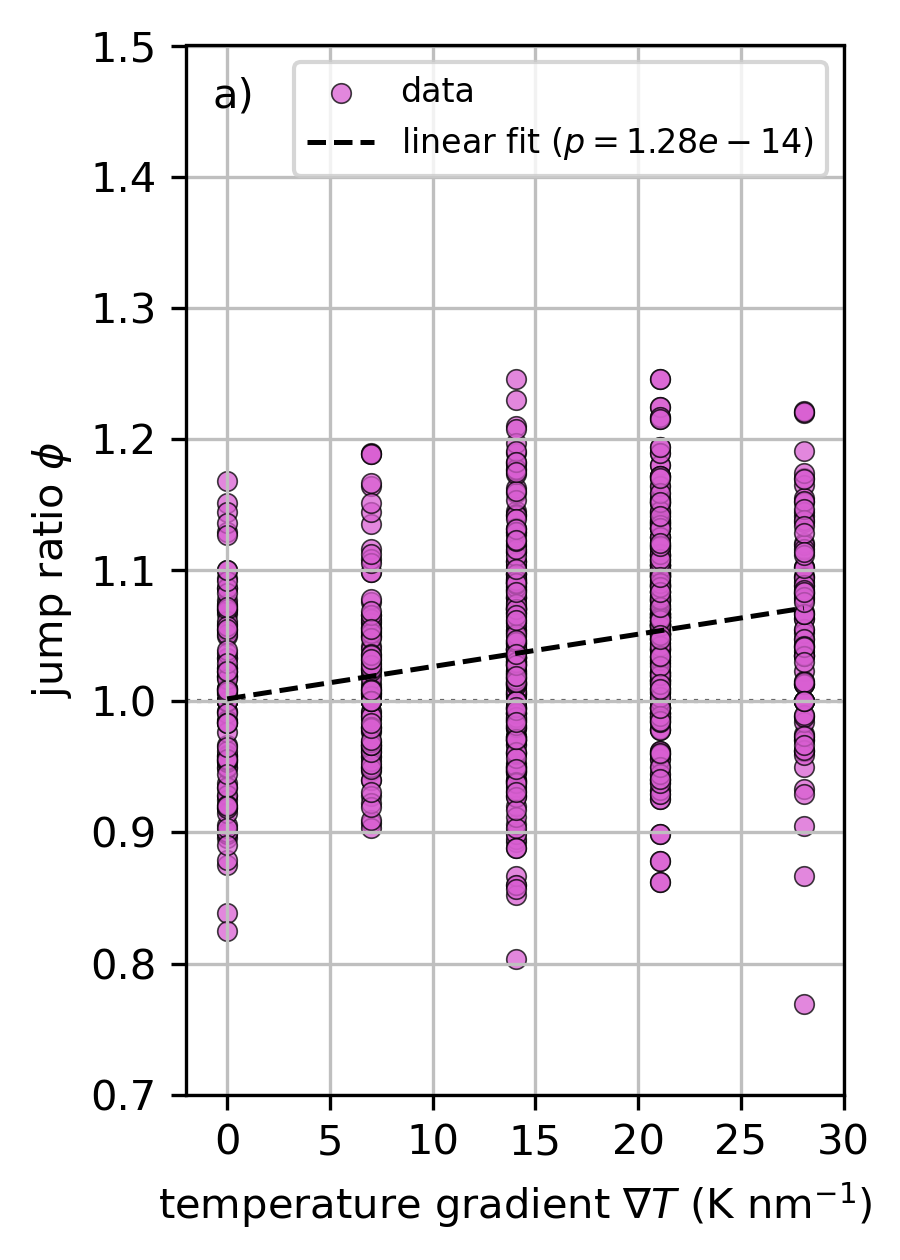}
        \vspace{-0.15cm}

       {\footnotesize\textbf{(a)} Vacancy}
    \end{minipage}
    \hspace{0.045\textwidth}
    \begin{minipage}[t]{0.27\textwidth}
        \centering
        \includegraphics[
            width=0.94\linewidth,
            trim=0.15cm 0.3cm 0.15cm 0.3cm,
            clip
        ]{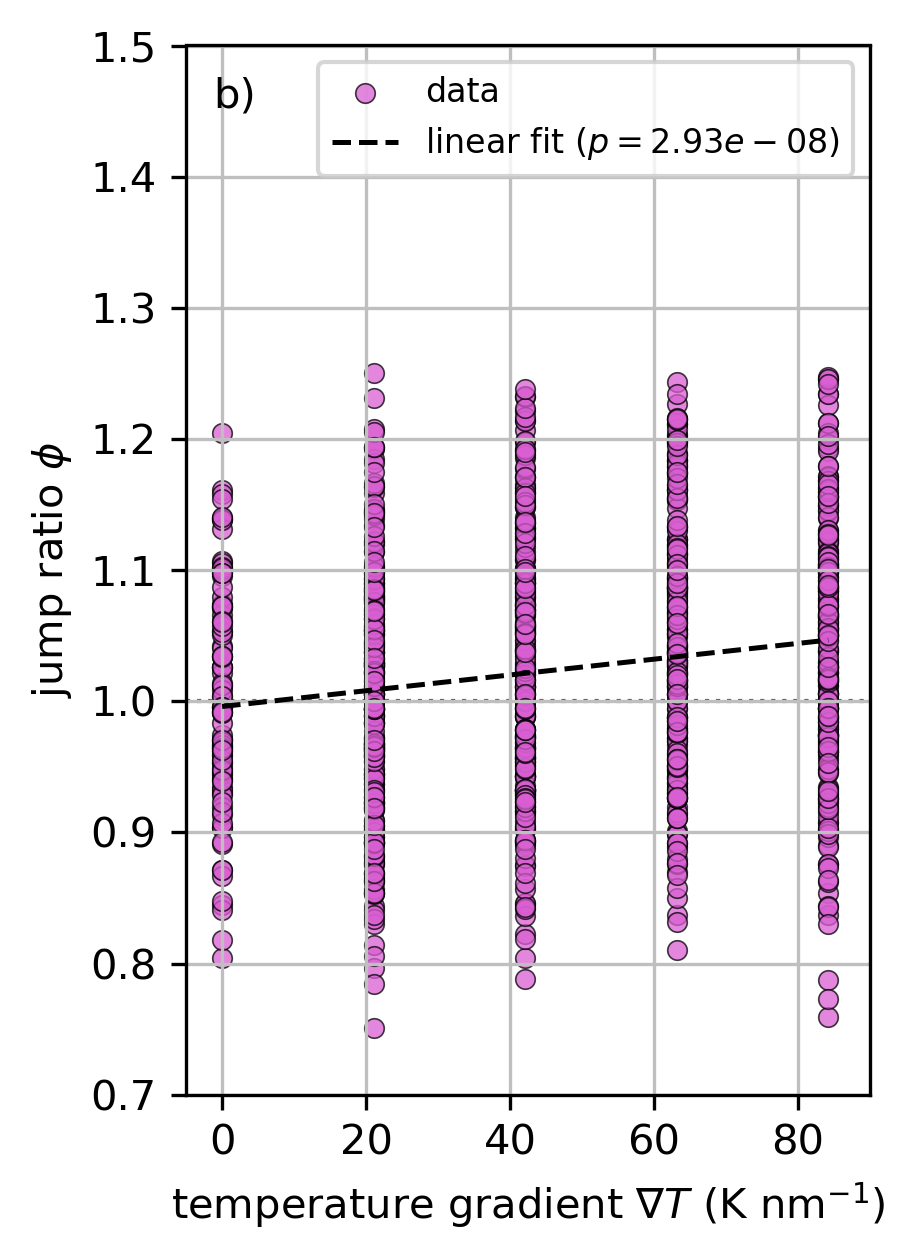}
        \vspace{-0.15cm}

        {\footnotesize\textbf{(b)} He$_1$}
    \end{minipage}
    \hspace{0.045\textwidth}
    \begin{minipage}[t]{0.27\textwidth}
        \centering
        \includegraphics[
            width=0.94\linewidth,
            trim=0.15cm 0.3cm 0.15cm 0.3cm,
            clip
        ]{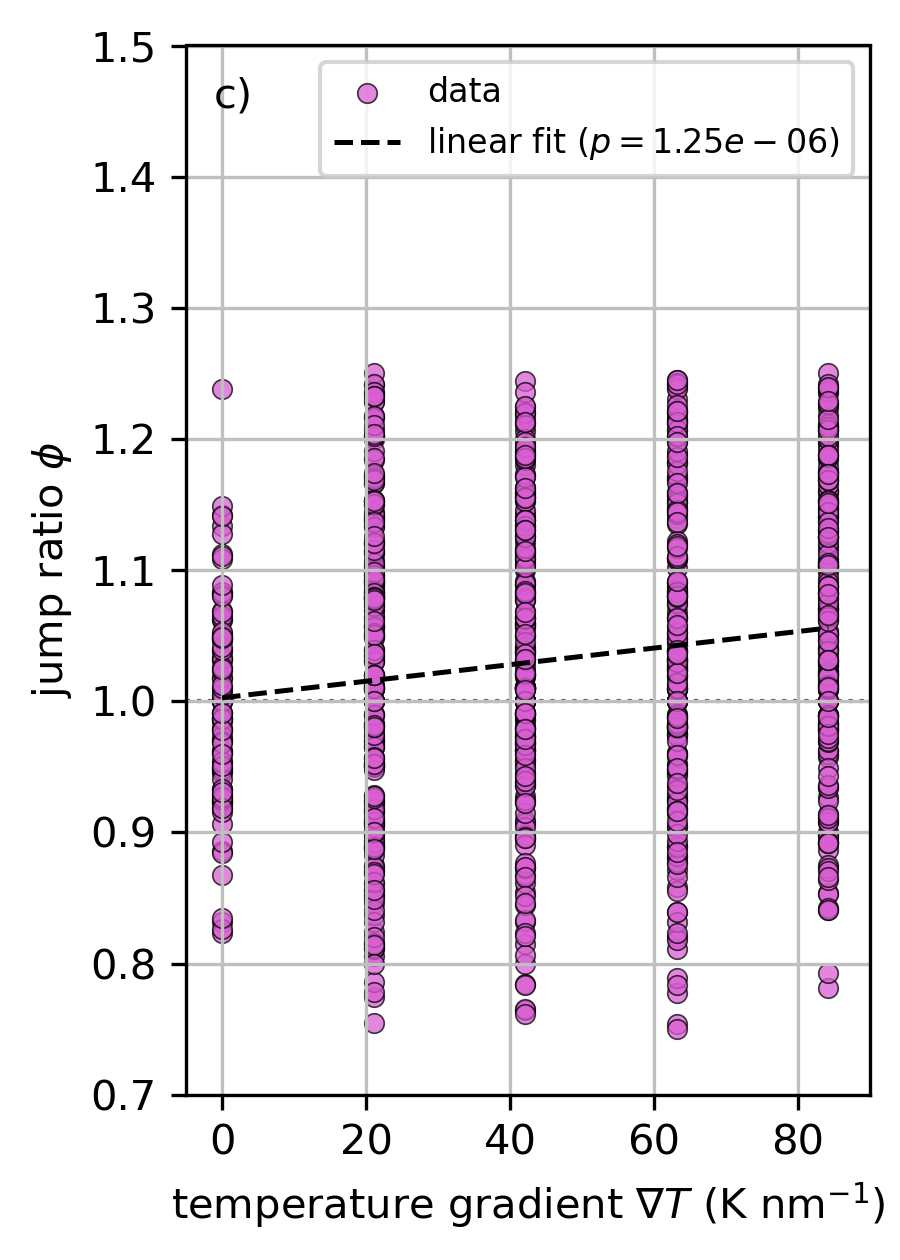}
        \vspace{-0.15cm}

        {\footnotesize\textbf{(c)} He$_2$}
    \end{minipage}

    \par\vspace{0.10cm}


    \begin{minipage}[t]{0.27\textwidth}
        \centering
        \includegraphics[
            width=0.94\linewidth,
            trim=0.15cm 0.3cm 0.15cm 0.3cm,
            clip
        ]{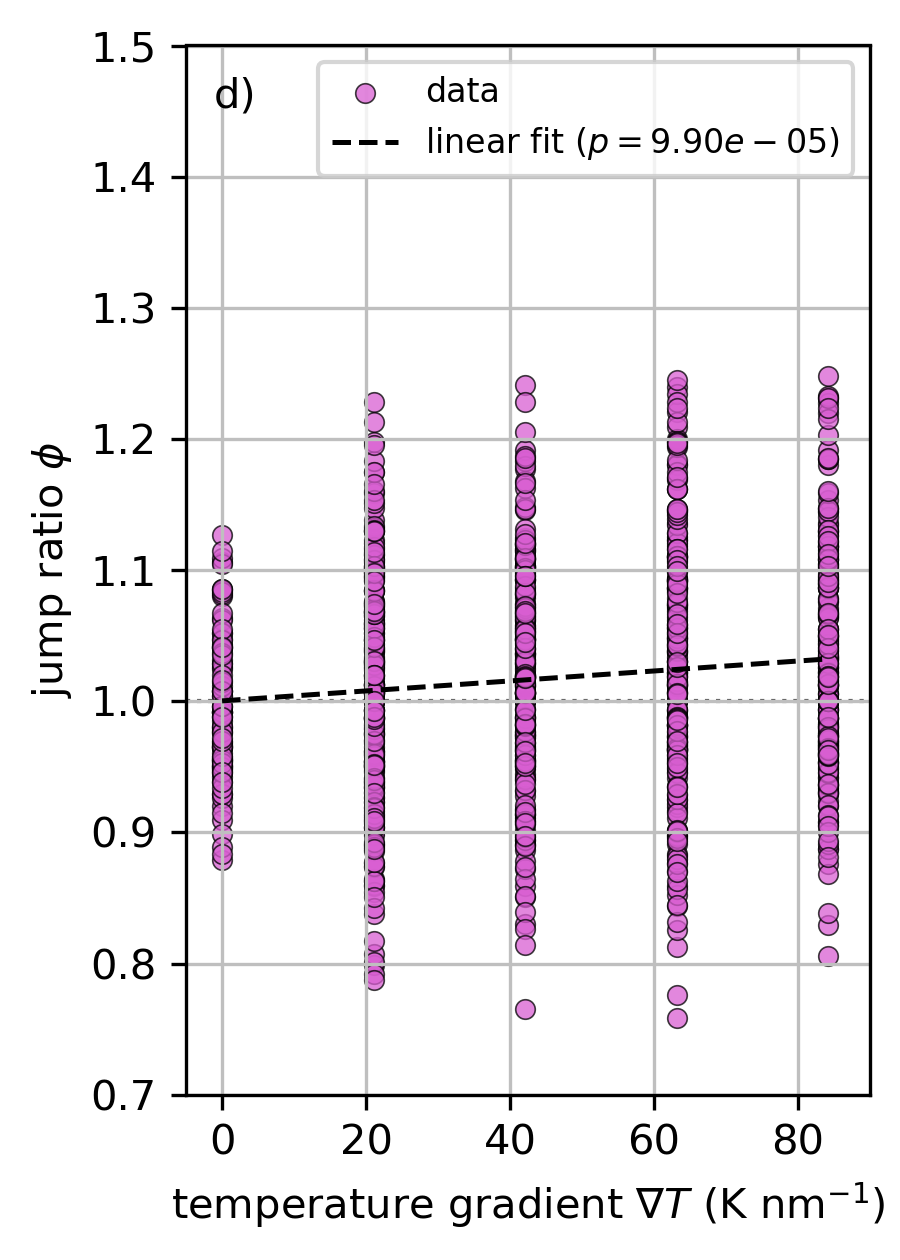}
        \vspace{-0.15cm}

        {\footnotesize\textbf{(d)} He$_3$}
    \end{minipage}
    \hspace{0.045\textwidth}
    \begin{minipage}[t]{0.27\textwidth}
        \centering
        \includegraphics[
            width=0.94\linewidth,
            trim=0.15cm 0.3cm 0.15cm 0.3cm,
            clip
        ]{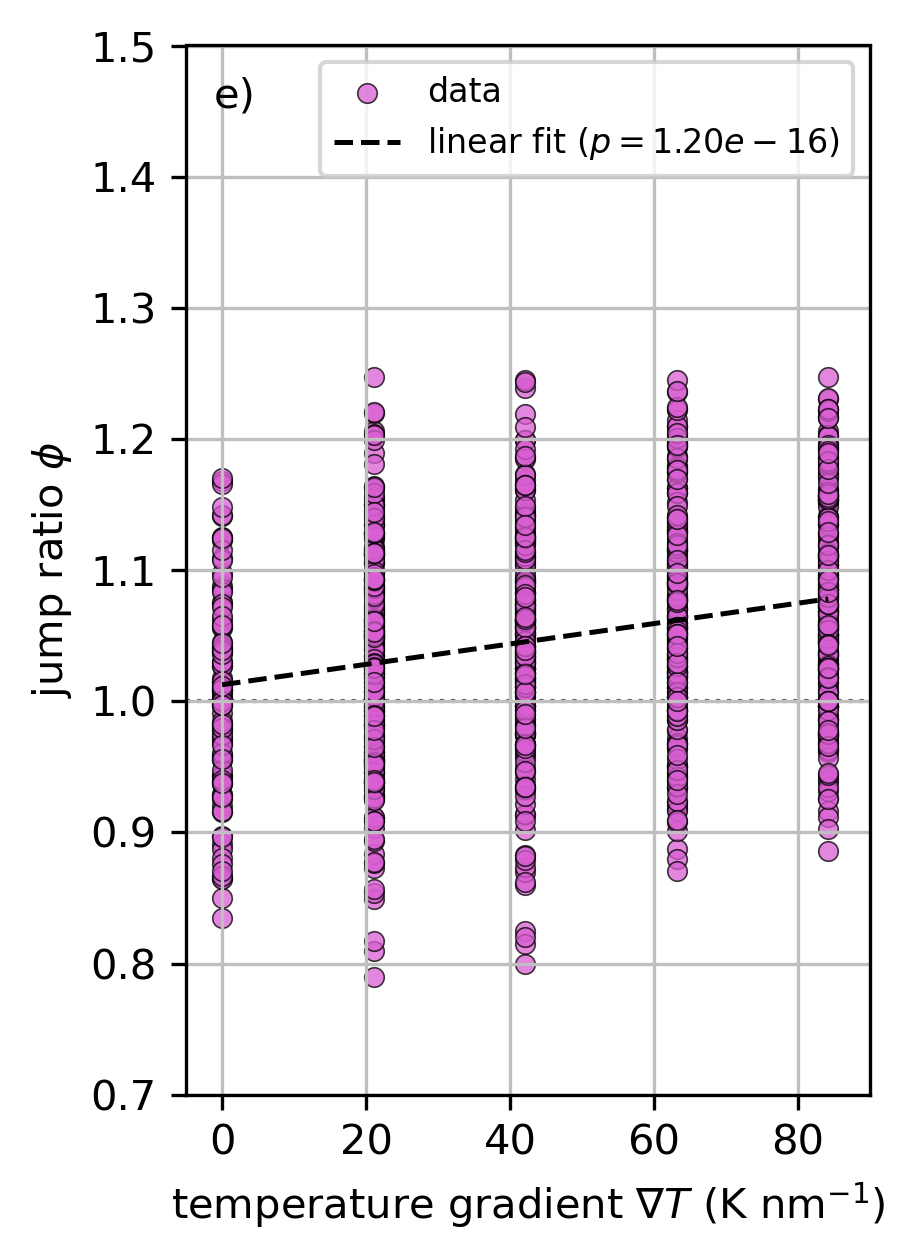}
        \vspace{-0.15cm}

        {\footnotesize\textbf{(e)} He$_4$}
    \end{minipage}

    \vspace{-0.20cm}

    \caption{{Directional jump ratio,
    $\phi=N_{\rm hot}/N_{\rm cold}$,
    as a function of the applied thermal gradient for (a) a single vacancy,
    (b) He$_1$, (c) He$_2$, (d) He$_3$, and (e) He$_4$.
    Each symbol represents one trajectory. }}
    \label{fig:phi_gradient}
\end{figure*}

\subsection{Helium-cluster diffusivity analysis}

Separate constant-temperature MD simulations were performed for
He$_1$--He$_4$ at 1000, 1300, and 1600~K using 100 statistically
independent trajectories for each cluster size and temperature. These
calculations were used only to characterize the equilibrium mobility of
the He clusters; vacancy diffusivity was not evaluated in the
present analysis. For a cluster containing $N_{\mathrm{He}}$ atoms, its
position was represented by the center of mass

\begin{equation}
\mathbf{R}_{\mathrm{COM}}(t)
=
\frac{1}{N_{\mathrm{He}}}
\sum_{i=1}^{N_{\mathrm{He}}}\mathbf{r}_i(t).
\label{eq:he_com}
\end{equation}

For He$_2$--He$_4$, displacements were included only while all He
atoms remained connected as a single cluster. If a cluster dissociated and later
reformed, a new trajectory segment was initiated, and no
displacement was evaluated across the dissociation interval. Using
multiple time origins from all valid segments, the three-dimensional
mean-squared displacement was calculated as

\begin{equation}
\mathrm{MSD}_{3\mathrm{D}}(\tau)
=
\left\langle
\left|\mathbf{R}_{\mathrm{COM}}(t+\tau)
-\mathbf{R}_{\mathrm{COM}}(t)\right|^2
\right\rangle,
\end{equation}

and the diffusivity was obtained from

{
\begin{equation}
D_{3\mathrm{D}}
=
\frac{1}{6}\frac{d\,\mathrm{MSD}_{3\mathrm{D}}}{d\tau}.
\label{eq:d3d}
\end{equation}
}

The MSD regression was weighted by the number of valid displacement
pairs at each lag time. A fitting interval of 20--420~ps was used for
all cases except He$_2$ at 1300 and 1600~K, for which the upper limits
were reduced to 240 and 150~ps, respectively, because dissociation
limited the available continuous trajectory lengths. Uncertainties in
$D_{3\mathrm{D}}$ and in the fitted migration energies were estimated
by bootstrap resampling of the 100 independent trajectories.

\section{Results}

Our previous investigation of hydrogen thermomigration in W established that the directional jump bias is governed primarily by the applied thermal gradient, whereas varying the average temperature at a fixed gradient produced no statistically significant change in the measured jump ratio~\cite{Alturk2026}. Accordingly, the thermomigration simulations in the present study focus on the influence of the thermal gradient while maintaining a fixed average temperature for each defect class. Separate constant-temperature simulations are then used to examine the equilibrium diffusivity and effective migration-energy scale of the He clusters.

\subsection{Directional jump statistics}

The influence of the applied thermal gradient on the migration bias was examined using the directional jump ratios obtained from all accepted independent simulations. Figure~\ref{fig:phi_gradient} presents the results for a single vacancy and He$_1$--He$_4$ as a function of the applied thermal gradient. The zero-gradient results provide an equilibrium check, for which $\phi$ is expected to be statistically consistent with unity. Linear least-squares fits with a free intercept are included to quantify the dependence of the directional jump ratio on the imposed thermal gradient for each defect species.

The trajectory-level ratios exhibit appreciable scatter,
as expected for thermally activated migration. Nevertheless, the fitted
slope is positive for the vacancy and for every He-cluster size,
indicating that increasing the thermal-gradient magnitude increases the
relative frequency of jumps toward the hot region.

The fitted slopes are statistically significant for every
defect species ($p<0.01$ in all cases). Thus, despite the variability
among individual trajectories, the jump statistics resolve a
measurable hot-directed thermomigration bias for both vacancies and
He clusters.

\subsection{Heat of transport}

\begin{table*}[t]
    \centering
    \caption{Characteristic projected jump lengths $\bar b$,
    directional-bias coefficients $\beta_{\phi}$, heat-of-transport
    coefficients $\alpha$, corresponding heat-of-transport expressions
    $Q^*(T)=\alpha kT^2$, and effective migration energies
    $E_m^{\mathrm{eff}}$ for a vacancy and He$_1$--He$_4$ in BCC tungsten.
    Reported uncertainties in $\beta_{\phi}$ and $\alpha$ are standard
    errors; parentheses give 95\% bootstrap confidence intervals for
    $E_m^{\mathrm{eff}}$.}
    \label{tab:heat_transport}

    \scriptsize
    \setlength{\tabcolsep}{3.5pt}
    \begin{tabular}{cccccc}
        \hline
        Defect &
        $\bar b$ (nm) &
        $\beta_{\phi}$ (nm\,K$^{-1}$) &
        $\alpha$ (K$^{-1}$) &
        $Q^*(T)=\alpha kT^2$ &
        $E_m^{\mathrm{eff}}$ (eV) \\
        \hline

        Vacancy &
        0.15826 &
        $(2.4658 \pm 0.3141)\times10^{-3}$ &
        $(-1.5581 \pm 0.1985)\times10^{-2}$ &
        $-1.5581\times10^{-2}kT^2$ &
        -- \\

        He$_1$ &
        0.07913 &
        $(6.0154 \pm 1.0766)\times10^{-4}$ &
        $(-7.6020 \pm 1.3605)\times10^{-3}$ &
        $-7.6020\times10^{-3}kT^2$ &
        $0.181\;(0.128\text{--}0.237)$ \\

        He$_2$ &
        0.05935 &
        $(6.3202 \pm 1.2955)\times10^{-4}$ &
        $(-1.0649 \pm 0.2183)\times10^{-2}$ &
        $-1.0649\times10^{-2}kT^2$ &
        $0.194\;(0.097\text{--}0.277)$ \\

        He$_3$ &
        0.05275 &
        $(3.7995 \pm 0.9722)\times10^{-4}$ &
        $(-7.2029 \pm 1.8430)\times10^{-3}$ &
        $-7.2029\times10^{-3}kT^2$ &
        $0.271\;(0.205\text{--}0.337)$ \\

        He$_4$ &
        0.03462 &
        $(7.7745 \pm 0.9228)\times10^{-4}$ &
        $(-2.2457 \pm 0.2666)\times10^{-2}$ &
        $-2.2457\times10^{-2}kT^2$ &
        $0.087\;(0.032\text{--}0.159)$ \\

        \hline
    \end{tabular}
\end{table*}
The measured dependence of the directional jump ratio on the applied thermal gradient enables the determination of the heat of transport directly from the jump-statistics framework. In our previous study of hydrogen thermomigration in W, the directional jump ratio was found to vary approximately linearly with the applied thermal gradient and was described by the first-order relation~\cite{Alturk2026}

\begin{equation}
\phi(|\nabla T|)=\phi_0+\beta_{\phi}|\nabla T|,
\label{eq:betaphi}
\end{equation}

\noindent where $\phi_0$ is the fitted zero-gradient intercept and
$\beta_{\phi}$ represents the sensitivity of the directional jump ratio
to the applied thermal-gradient magnitude. The fitted values of
$\phi_0$ are consistent with the equilibrium expectation $\phi_0=1$.
In the linear-response limit, substituting the fitted slope into
Eq.~(\ref{eq:qstar}) gives

\begin{equation}
Q^*(z)
=
-\frac{\beta_{\phi}}{\bar b}kT^2(z)
=
\alpha kT^2(z),
\end{equation}

\noindent with

\begin{equation}
\alpha=-\frac{\beta_{\phi}}{\bar b}.
\end{equation}

Table~\ref{tab:heat_transport} summarizes the characteristic jump
lengths, fitted values of $\beta_{\phi}$, and the resulting heat-of-transport
coefficient $\alpha$ for the vacancy and He$_1$--He$_4$ defects. Here,
$\bar b$ is defined as the mean absolute projection of an elementary
defect displacement along the temperature-gradient direction,
$z\parallel[001]$. For the vacancy, migration occurs through a
nearest-neighbor atom--vacancy exchange in the BCC lattice. Each of the
eight displacement vectors, $(a/2)(\pm1,\pm1,\pm1)$, has
an absolute $z$ projection of $a/2$. Therefore,
$\bar b_{\mathrm V}=a/2=0.15826$~nm, using
$a=0.31652$~nm as the lattice parameter for W.

\begin{figure*}[!t]
    \centering
    \includegraphics[width=0.90\textwidth]
    {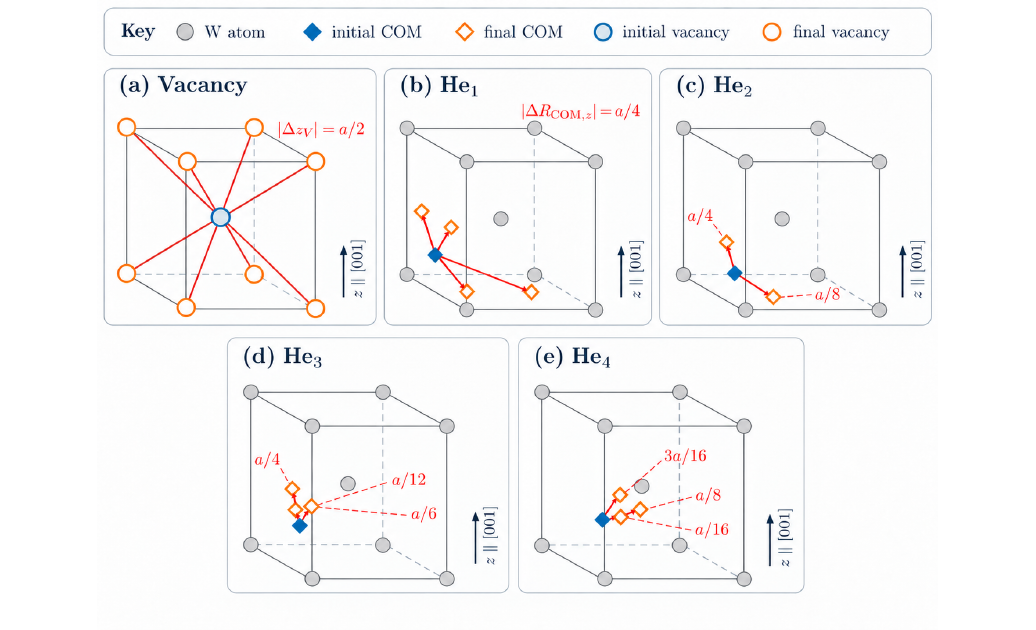}
    \caption{Representative jump geometries used to determine the
    characteristic projected jump lengths for a vacancy and
    He$_1$--He$_4$ in BCC tungsten.}
    \label{fig:defect_jump_geometry}
\end{figure*}

For the He defects, the jump lengths were determined geometrically
from nearest-neighbor transitions between tetrahedral interstitial
sites. For a cluster containing $N$ He atoms, the projected
center-of-mass displacement associated with an admissible transition
$j$ was calculated as
\begin{equation}
b_{N,j}
=
\left|
\Delta R_{\mathrm{COM},z}^{(j)}
\right|
=
\left|
\frac{1}{N}
\sum_{i=1}^{N}
\left(z_{i,f}^{(j)}-z_{i,i}^{(j)}\right)
\right|.
\end{equation}
All admissible geometrical transitions were assigned equal probability,
and transitions producing the same value of $b_{N,j}$ were grouped
according to their degeneracy $g_j$. The characteristic jump length was
then evaluated as the probability-weighted average
\begin{equation}
\bar b_N
=
\sum_j P_j b_{N,j}
=
\frac{\sum_j g_j b_{N,j}}{\sum_j g_j},
\qquad
P_j=\frac{g_j}{\sum_k g_k}.
\label{eq:b_average}
\end{equation}
The admissible transitions were required to involve
nearest-neighbor tetrahedral-site hops, preserve distinct He
occupations and the compact cluster geometry, and produce a nonzero
center-of-mass displacement. The distinct geometrical jump classes used
to evaluate $b_{N,j}$ are
illustrated in Fig.~\ref{fig:defect_jump_geometry}. For the vacancy, the
marked positions denote vacant W lattice sites; for the He
defects, they denote the cluster center-of-mass. The arrows show the
possible defect displacements, and their labels give the absolute
projection along $z\parallel[001]$. For He$_2$--He$_4$, one
representative of each distinct projected jump length is shown, with
symmetry-equivalent transitions included through the degeneracies
$g_j$ in Eq.~(\ref{eq:b_average}).

For He$_1$, all transitions have $b=a/4$, giving
$\bar b_1=a/4$. For He$_2$, the possible magnitudes
$a/8$ and $a/4$ each have probability
$1/2$, giving $\bar b_2=3a/16$. For He$_3$, the
magnitudes $a/12$, $a/6$, and
$a/4$ have probabilities $1/4$, $1/2$, and $1/4$,
respectively, giving $\bar b_3=a/6$. For He$_4$, the
magnitudes $a/16$, $a/8$, and
$3a/16$ have probabilities $3/8$, $1/2$, and $1/8$,
respectively, giving $\bar b_4=7a/64$. Using the W lattice parameter discussed above produces the He jump lengths reported
in Table~\ref{tab:heat_transport}. Expressing the heat-of-transport as
$Q^*=\alpha kT^2$ separates the intrinsic defect-dependent coefficient
from the explicit temperature dependence, thereby allowing direct
comparison between vacancy and He-cluster simulations performed at
different temperatures.

Since all fitted values of $\beta_{\phi}$ are positive, the corresponding heat-of-transport coefficients are negative, indicating that every investigated defect exhibits a net thermomigration tendency toward the hot region. Although the migration direction is identical for all defect species, the magnitude of the response varies considerably. He$_4$ displays the largest magnitude of $\alpha$, followed closely by the vacancy, whereas He$_3$ exhibits the smallest value. This variation reflects the combined influence of the measured directional jump bias and the characteristic migration distance.

Overall, these results demonstrate that the jump-statistics framework developed previously for hydrogen transport can be successfully extended to fundamentally different defect species. Despite the distinct migration mechanisms associated with vacancies and He clusters, the same statistical framework provides a consistent and quantitative measure of their thermomigration behavior through the heat-of-transport.

\subsection{Helium-cluster diffusivity and effective migration energies}

To examine whether the anomalous thermomigration response of He$_3$ is
also reflected in its equilibrium mobility, the three-dimensional
diffusivities of He$_1$--He$_4$ were evaluated at 1000, 1300, and
1600~K. The temperature dependence was described by

\begin{equation}
D_{3\mathrm{D}}(T)
=
D_0\exp\left(-\frac{E_m^{\mathrm{eff}}}{k_{\mathrm B}T}\right).
\label{eq:arrhenius}
\end{equation}

Defining the inverse thermal energy as
$\beta_T=1/(k_{\mathrm B}T)$ gives

\begin{equation}
\ln D_{3\mathrm{D}}
=
\ln D_0-E_m^{\mathrm{eff}}\beta_T,
\end{equation}

so that the slope of $\ln D_{3\mathrm{D}}$ versus $\beta_T$ is
$-E_m^{\mathrm{eff}}$. This inverse-temperature variable is distinct
from the directional-bias coefficient $\beta_{\phi}$ used in the
thermomigration analysis.

\begin{figure*}[!t]
    \centering
    \includegraphics[width=0.75\textwidth]
    {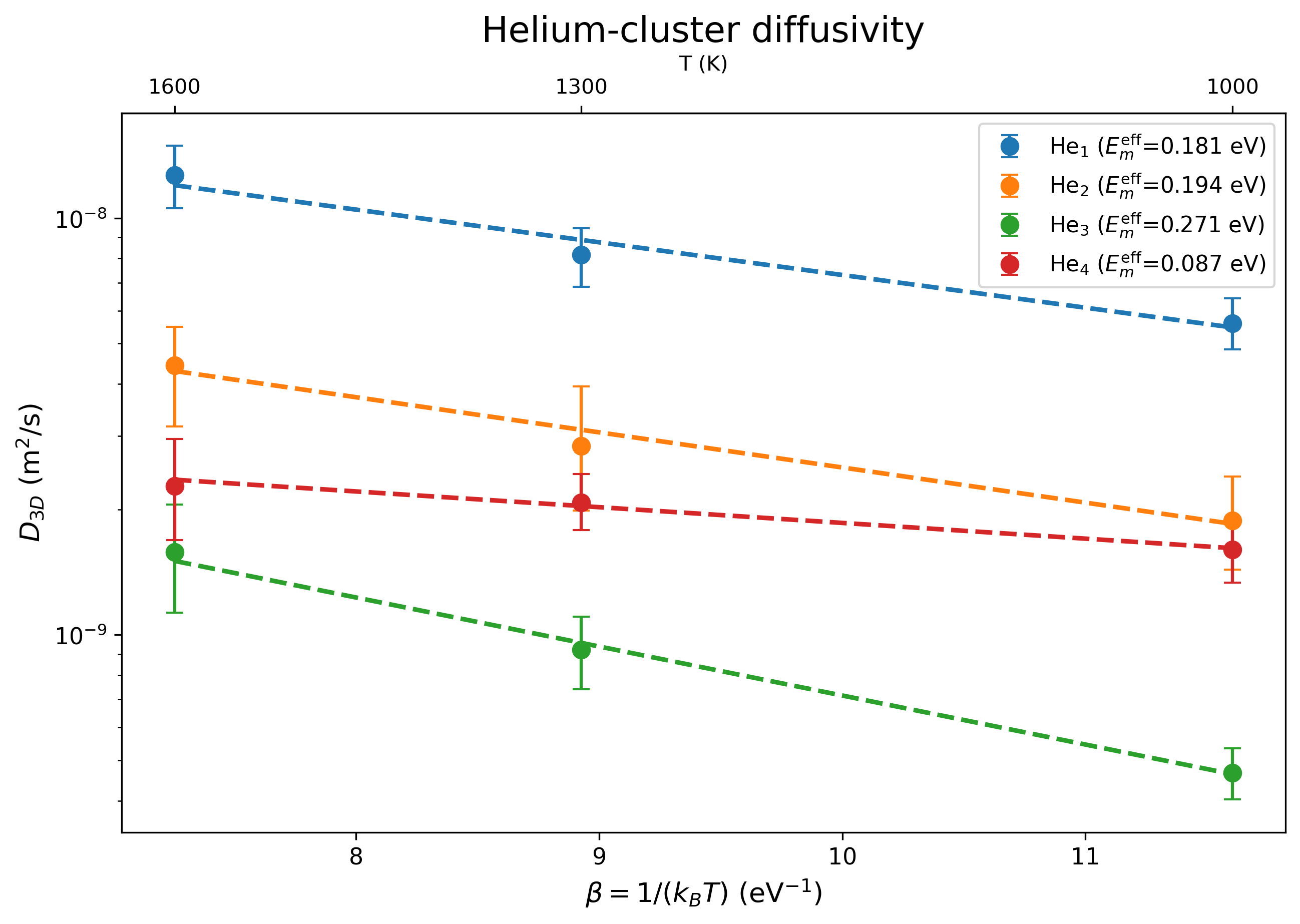}
    \caption{Three-dimensional diffusivity of He$_1$--He$_4$ as a
    function of inverse thermal energy,
    $\beta_T=1/(k_{\mathrm B}T)$, shown on a logarithmic scale. The
    horizontal-axis symbol $\beta$ denotes $\beta_T$ and is distinct
    from the directional-bias coefficient $\beta_{\phi}$. Symbols are
    the MD results, error bars show bootstrap uncertainty, and dashed
    lines are Arrhenius fits over 1000--1600~K.}
    \label{fig:he_diffusivity_beta}
\end{figure*}

At all three temperatures, the mobility follows

\begin{equation}
D_{\mathrm{He}_1}
>
D_{\mathrm{He}_2}
>
D_{\mathrm{He}_4}
>
D_{\mathrm{He}_3}.
\end{equation}

Thus, diffusivity does not decrease monotonically with cluster size:
He$_3$ is consistently the least mobile cluster throughout the investigated temperature
range. The Arrhenius fits in Fig.~\ref{fig:he_diffusivity_beta} give
$E_m^{\mathrm{eff}}=0.181$, 0.194, 0.271, and 0.087~eV for
He$_1$--He$_4$, respectively, with the confidence intervals listed in
Table~\ref{tab:heat_transport}. The resulting ordering is

\begin{equation}
E_m^{\mathrm{eff}}(\mathrm{He}_3)
>
E_m^{\mathrm{eff}}(\mathrm{He}_2)
\simeq
E_m^{\mathrm{eff}}(\mathrm{He}_1)
>
E_m^{\mathrm{eff}}(\mathrm{He}_4).
\end{equation}

He$_2$ becomes increasingly unstable as temperature increases. At
1600~K, approximately 76\% of the sampled configurations remain
compact, but the intact intervals are divided among many short
segments. Its reported diffusivity therefore describes the motion of
the He$_2$ pair during intervals in which the two atoms remain
bound.

\section{Discussion}

The present work demonstrates that the jump-statistics framework
previously developed for hydrogen thermomigration in W is not
specific to hydrogen interstitial diffusion, but can be extended to
fundamentally different defect species. Although hydrogen atoms,
vacancies, and He clusters migrate through distinct atomistic
mechanisms, the same statistical description based on directional jump
probabilities provides a consistent means of determining the heat-of-transport directly from NEMD trajectories. This
generality is significant because it avoids the need to determine
steady-state concentration profiles, which can require prohibitively
long trajectories for slowly diffusing defects or complex defect
clusters. Instead, the thermomigration response is extracted directly
from the statistics of individual migration events, providing a unified
framework applicable to a broad range of defect species.

As an independent validation of the jump-statistics approach, the
steady-state concentration distribution predicted from the measured
directional jump ratio was compared with that obtained directly from a
long He$_3$ trajectory. Following the jump-based formulation developed
previously for hydrogen thermomigration in W~\cite{Alturk2026},
the zero-flux condition gives

\begin{equation}
\frac{dc}{dz}
=
-\frac{2}{\bar b}
\left(\frac{1-\phi}{1+\phi}\right)c.
\label{eq:concentration_gradient}
\end{equation}

For a constant directional jump ratio over a region in which the
thermal-gradient direction is fixed, integration gives

\begin{equation}
c(z)
=
c(z_0)
\exp\left[
-\frac{2(1-\phi)}
{\bar b(1+\phi)}
(z-z_0)
\right].
\label{eq:concentration_profile}
\end{equation}

The profile was evaluated using the independently determined He$_3$
directional jump ratio and characteristic jump length, with the
normalization fixed by the average concentration in the simulation
cell. Because the hot region is located at the center of the simulation
domain, Eq.~(\ref{eq:concentration_profile}) was applied separately on
either side of the center, where the direction toward the hot region
reverses.

\begin{figure*}[!t]
    \centering
    \includegraphics[width=0.60\textwidth]
    {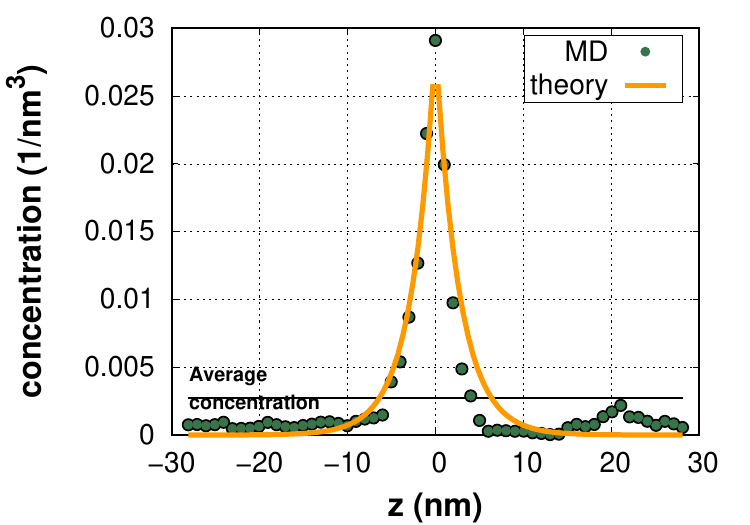}
    \caption{Steady-state He$_3$ concentration profile obtained from a
    long molecular-dynamics trajectory compared with the profile
    predicted from the independently determined directional jump ratio
    $\phi$ (solid line).}
    \label{fig:He3_profile}
\end{figure*}

As shown in Fig.~\ref{fig:He3_profile}, both approaches predict
pronounced accumulation of He$_3$ in the hot region. The profile
calculated from the independently measured jump statistics reproduces
the location and overall shape of the concentration peak obtained from
the long NEMD trajectory. The remaining variations in the
long-trajectory concentration profile are consistent with the finite
sampling associated with following a single diffusing cluster over a
finite simulation time. Importantly, the predicted profile is obtained
from the directional jump statistics rather than by fitting the
long-trajectory concentration data. The agreement therefore provides an
independent validation of the jump-statistics methodology and shows that
an ensemble of shorter, statistically independent simulations can
recover the thermomigration behavior observed directly over a much
longer simulation timescale.

The He$_1$ result is also consistent with the earlier NEMD study of
Martinez \textit{et al}, who obtained
\mbox{$Q^*=-6.5\times10^{-3}kT^2$~}  for a single interstitial He atom
by fitting its steady-state concentration profile
\cite{Martinez2021}. The present jump-statistics result,
$Q^*=(-7.6020\pm1.3605)\times10^{-3}kT^2$~, agrees with the
previous value within the uncertainty of the fitted coefficient.
This agreement is encouraging because the two values
were obtained from different observables and provides further support
for the directional jump-statistics approach.

The negative values of $Q^*$ obtained for all investigated defect
species have important implications for defect redistribution in
plasma-facing W. Within the present sign convention, a negative
heat-of-transport corresponds to thermomigration toward regions of
higher temperature. During plasma exposure, for which the
plasma-facing surface is generally hotter than the underlying bulk,
thermomigration therefore provides a driving force for both vacancies
and He clusters toward the surface. This contribution can modify
the defect concentration profiles that would otherwise result from
Fickian diffusion, He implantation, irradiation-induced defect
production, trapping, and interactions with microstructural sinks.
The simultaneous hot-side migration of vacancies and He clusters
may also increase their spatial overlap in the near-surface region,
potentially affecting He--vacancy association, He retention, bubble
nucleation and growth, and subsequent surface evolution. The magnitude
of these effects will ultimately depend on the competition between
thermomigration and the other transport, reaction, and sink mechanisms
operating under irradiation conditions.

Although the direction of thermomigration is the same for all defects
considered here, its magnitude is strongly species-dependent. He$_4$
exhibits the largest magnitude of $\alpha$, followed by the
vacancy, whereas He$_3$ exhibits the smallest response. Interestingly,
the relatively large responses obtained for He$_4$ and the vacancy
arise from different combinations of the quantities entering
$\alpha=-\beta_{\phi}/\bar b$. The directional-bias coefficient of
the vacancy is substantially larger than that of He$_4$, but its
characteristic jump length is also much larger. For He$_4$, the
smaller directional bias is offset by its considerably shorter
characteristic migration distance. This illustrates that the heat-of-transport cannot
be inferred from the directional bias or migration geometry alone, but
is determined by their combined effect.

For the He clusters, the characteristic jump length decreases
progressively from He$_1$ to He$_4$. Because the fitted values of
$\beta_{\phi}$ for He$_1$, He$_2$, and He$_4$ are similar, the decreasing
$\bar b$ produces a general increase in $|\alpha|$ with cluster size.
He$_3$, however, deviates strongly from this trend. Its characteristic
jump length lies between those of He$_2$ and He$_4$, but its fitted
directional-bias coefficient is only approximately 60\% of the He$_2$
value and 49\% of the He$_4$ value. The anomalously small
$|\alpha|$ of He$_3$ therefore originates primarily from its weak
directional response to the applied thermal gradient rather than from
an anomalous characteristic migration distance.

The constant-temperature calculations provide an independent indication
that He$_3$ has a distinct migration landscape. He$_3$ is the least
mobile cluster at all three temperatures and has the largest effective
migration energy, $E_m^{\mathrm{eff}}=0.271$~eV. By comparison,
He$_1$, He$_2$, and He$_4$ have effective migration energies of 0.181,
0.194, and 0.087~eV, respectively. Thus, the weak directional response
of He$_3$ coincides with reduced equilibrium mobility and a steeper
temperature dependence of its diffusivity. This correspondence does
not imply that the migration energy alone determines $Q^*$; rather, it
supports the interpretation that the anomalous He$_3$ thermomigration
response is associated with its defect-specific migration landscape.

This behavior is consistent with the distinct structure and migration
landscape reported for small interstitial He clusters in W.
Perez \textit{et al.} showed that the minimum-energy migration pathway of
He$_3$ contains a relatively deep intermediate metastable minimum that is
nearly degenerate with the ground-state configuration
\cite{Perez2014}. Such an energy landscape permits sampling of multiple
closely related configurations during migration, including reversible
transitions that do not necessarily result in complete net translation.
Moreover, the equilibrium, metastable, and transition-state
configurations possess different vibrational spectra and may therefore
couple differently to the nonequilibrium phonon population generated by
the thermal gradient.
 Such defect-specific phonon coupling provides a
possible microscopic origin for the reduced directional-bias
coefficient observed for He$_3$. A direct mode-resolved analysis would
be required to establish this relationship quantitatively; nevertheless,
the present results suggest that He-cluster thermomigration depends
not simply on cluster size or migration barrier, but also on the
defect-specific migration landscape and its coupling to the
heat-carrying vibrational modes.

The migration energies extracted here are effective high-temperature
activation energies obtained from the MD diffusivity over
1000--1600~K, rather than individual zero-temperature saddle-point
barriers. Comparison with the zero-temperature barriers reported by
Perez \textit{et al.} is qualitative because their calculations used
the Ackland--Thetford W--W potential modified by Juslin and
Wirth~\cite{Juslin2013,Ackland1987}, whereas the present simulations
use the Marinica \textit{et al.} W--W potential ~\cite{Marinica2013}.

Perez \textit{et al.} also showed that small helium clusters in
tungsten can exhibit non-Arrhenius diffusion because thermally
accessible internal configurations and migration pathways change with
temperature~\cite{Perez2014}. The present values may therefore include
entropic contributions and the combined effects of multiple migration
mechanisms, and differences from static nudged-elastic band (NEB) barriers should not be
interpreted as inconsistencies.

Migration energy and heat-of-transport characterize different aspects
of defect motion. In Schottky's lattice-dynamical treatment \cite{Schottky1963}, their
relationship for a one-dimensional vacancy model can be written in the
form

\begin{equation}
\frac{Q^*}{E_m}
=
-\frac{2\tau w_1}{\bar{b}},
\label{eq:schottky_ratio}
\end{equation}

where $\bar{b}$ is the jump distance, $\tau$ is the relaxation time of the
nonequilibrium phonon distribution, and $w_1$ describes the dynamical
response of the lattice~\cite{Schottky1963}. The migration energy thus
provides an energetic scale, but the ratio $Q^*/E_m$ also depends on
the migration geometry and coupling to the lattice vibrations. The
ratios obtained at 1300~K are summarized in
Table~\ref{tab:qstar_em_jump}. Their ordering does not follow the
ordering of $E_m^{\mathrm{eff}}$, confirming that the migration energy
alone does not determine the thermomigration response.

\begin{table}[!tbp]
    \centering
    \caption{Heat-of-transport-to-migration-energy ratios for helium
    clusters at 1300~K.}
    \label{tab:qstar_em_jump}
    \resizebox{\columnwidth}{!}{%
    \begin{tabular}{cccc}
        \toprule
        Cluster &
        $\bar b$ (nm) &
        $Q^*/E_m^{\mathrm{eff}}$ &
        $\bar b(Q^*/E_m^{\mathrm{eff}})$ (nm) \\
        \midrule
        He$_1$ & 0.07913 & $-6.1$  & $-0.48$ \\
        He$_2$ & 0.05935 & $-8.0$  & $-0.47$ \\
        He$_3$ & 0.05275 & $-3.9$  & $-0.20$ \\
        He$_4$ & 0.03462 & $-38$ & $-1.3$ \\
        \bottomrule
    \end{tabular}%
    }
\end{table}

Multiplying $Q^*/E_m^{\mathrm{eff}}$ by the characteristic
projected center-of-mass jump length gives the length-scaled ratio in
the final column of Table~\ref{tab:qstar_em_jump}. He$_1$ and He$_2$
then show nearly identical values, while He$_3$ has a substantially
smaller magnitude and He$_4$ remains a strong outlier.
Because
Eq.~(\ref{eq:schottky_ratio}) was derived for a one-dimensional vacancy
model, these quantities are used only as a qualitative comparison and
not as a quantitative determination of lattice force constants or
phonon relaxation times for the He clusters.

The heat-of-transport coefficients obtained in this work provide
atomistically derived parameters for higher-length-scale descriptions
of defect evolution in W under thermal gradients. Continuum
approaches, including rate-theory and cluster-dynamics models with reaction, diffusion and advection terms, require
such parameters to account for thermally driven defect redistribution. Rate-theory studies have shown that defect diffusivities and effective migration energies strongly control vacancy--interstitial recombination and the resulting accumulation of radiation damage \cite{Nahavandian2025}.

The strong species dependence observed here further indicates that
thermomigration parameters should not be inferred solely from defect
size or diffusivity. Incorporating defect-specific heats-of-transport
into mesoscale models will allow the competition between thermal
gradients, diffusion, irradiation-induced production, trapping, and
defect reactions to be assessed and will help determine how
thermomigration modifies He and vacancy distributions in
plasma-facing W.

\section{Conclusions}\label{sec:conclusions}
The jump-statistics framework previously developed for hydrogen has been extended to quantify the thermomigration of a vacancy and He$_1$--He$_4$ clusters in BCC W. Ensembles of independent non-equilibrium molecular dynamics (NEMD) trajectories were used to measure the directional jump ratio $\beta _{\phi}$, defined in Eq.\ref{eq:betaphi}, as a function of the applied thermal gradient. The fitted zero-gradient intercepts were consistent with the equilibrium expectation $\phi_0=1$, while the positive slopes obtained for every defect demonstrate a systematic bias toward the hot region. Combining these slopes with defect-specific projected jump lengths yielded negative heat-of-transport coefficients for all five defect species.

The thermomigration magnitude is strongly species-dependent.
He$_4$ has the largest value of $|\alpha|$,
$2.2457\times10^{-2}$~K$^{-1}$, followed by the vacancy,
whereas He$_3$ has the smallest response. For He$_1$, He$_2$, and
He$_4$, similar directional-bias coefficients combined with
progressively shorter projected jump lengths produce a general increase
in $|\alpha|$ with cluster size. He$_3$ breaks this trend because its
directional-bias coefficient is smaller, rather than because of an
anomalous jump length.

The equilibrium diffusivity calculations independently show
that He$_3$ has a distinct migration landscape. It is the least mobile
cluster over 1000--1600~K and has the largest effective migration
energy, 0.271~eV, whereas He$_4$ is more mobile and has the smallest
effective migration energy, 0.087~eV. This correspondence supports the
interpretation that the weak thermomigration response of He$_3$ is
connected to its defect-specific migration landscape. Nevertheless,
the lack of a common ordering among $Q^*$, diffusivity, and
$E_m^{\mathrm{eff}}$, together with the variation in
$Q^*/E_m^{\mathrm{eff}}$, confirms that the migration barrier alone
does not determine the heat-of-transport.

Finally, the He$_3$ steady-state concentration profile predicted from the independently measured jump statistics agrees with that obtained from a much longer NEMD trajectory, providing a direct validation of the approach. The method therefore offers an efficient route for extracting defect-specific thermomigration parameters without requiring steady-state simulations for every species. The predicted hot-side migration of both vacancies and He clusters may enhance their near-surface redistribution and association in plasma-facing W, and the coefficients reported here can be incorporated into rate-theory, cluster-dynamics, and other mesoscale models of defect evolution under thermal gradients.

\section{Acknowledgments}\label{sec:acknowledgments}
Discussions with S. Blondel, D. Dasgupta, D. Maroudas, and
B. D. Wirth are gratefully acknowledged. This work was supported by the
U.S. Department of Energy, Office of Science, Established Program to
Stimulate Competitive Research (EPSCoR), under Award No.
DE-SC-0023385. Computational resources were provided by the Clemson
University Palmetto Cluster under National Science Foundation awards
MRI\#2024205, MRI\#1725573, and CRI\#2010270.

\bibliographystyle{model1-num-names}

\bibliography{references}


\end{document}